\documentclass[journal,draftcls,onecolumn,10pt,twoside]{IEEEtran}

\usepackage{amstext}
\usepackage{multirow}
\usepackage{graphicx}
\usepackage{amssymb}
\usepackage{amsmath}
\usepackage{algorithmic}
\usepackage{algorithm}
\usepackage[T1]{fontenc}
\usepackage[latin9]{inputenc}
\usepackage{float}
\usepackage{bigstrut}
\usepackage{cite}

\def\Pa{\mathbf{P{\hspace{-0.1cm}}_A}}
\def\Paa{\mathbf{P{\hspace{-0.1cm}}_{A_1}}}
\def\Pab{\mathbf{P{\hspace{-0.1cm}}_{A_2}}}

\ifCLASSINFOpdf
\else
\fi

\begin{document}
%
\title{Maximum Likelihood Algorithms for Joint Estimation of Synchronization
Impairments and Channel in MIMO-OFDM System}
%
%
%
\author{Renu Jose
                 and~K.V.S. Hari\\
SSP Lab, Department
of Electrical Communication Engineering, Indian Institute of Science,
Bangalore-560012, India. e-mail: renujose@ece.iisc.ernet.in,
}
\thanks{"This paper is a preprint of a paper submitted to IET Communications
and is subject to Institution of Engineering and Technology Copyright. If
accepted, the copy of record will be available at IET Digital Library"}
\maketitle

\begin{abstract}
Maximum Likelihood (ML) algorithms, for the joint estimation of synchronization
impairments and channel in Multiple Input Multiple Output-Orthogonal Frequency
Division Multiplexing (MIMO-OFDM) system, 
are investigated in this work. A system model that takes into account the
effects of
carrier frequency offset, sampling frequency offset, symbol timing error, and
channel impulse response is formulated. Cram\'{e}r-Rao Lower Bounds for
the estimation of continuous parameters are derived, which show the coupling
effect among different impairments and the significance of the joint
estimation. 
We propose an ML algorithm for the estimation of synchronization
impairments and channel together, using grid search method.
To reduce the complexity of the joint grid search in ML algorithm, a Modified ML
(MML) algorithm with multiple
one-dimensional searches is also proposed. Further, a
Stage-wise ML (SML) algorithm using existing algorithms, which estimate
fewer number of parameters, is also proposed.
Performance of the estimation algorithms is studied through numerical
simulations and it is found that the proposed ML and MML algorithms exhibit
better performance than SML algorithm.
\end{abstract}


\section{Introduction}
\IEEEPARstart{T}{he} integration of Multiple Input Multiple Output (MIMO) and
Orthogonal Frequency Division Multiplexing (OFDM) techniques has become a
preferred solution for 
the high rate wireless technologies due to its high spectral efficiency,
robustness to frequency selective fading,
increased diversity gain, and enhanced system capacity.
The main drawback of OFDM based systems is their susceptibility to
synchronization impairments such as Carrier Frequency Offset (CFO), Sampling
Frequency Offset (SFO) and Symbol Timing Error
(STE)~\cite{Meyr},~\cite{morelli_tutorial}.
\let\thefootnote\relax\footnotetext{{Notations:}
Upper case bold letters denote matrices and lower case bold letters denote
column vectors. 
$\hat{\mathbf{A}}$ denotes the estimate of $\mathbf{A}$.
 $\Re(\mathbf{A})$ and $\Im(\mathbf{A})$ denote the real and imaginary parts 
of the elements of $\mathbf{A}$, respectively.
$\mathbf{1}$ and $\mathbf{0}$ represent the all-one and all-zero column vector,
respectively.
 $\frac{\partial(\mathbf{A})}{\partial\epsilon}$ represents the partial
derivative of $\mathbf{A}$
with respect to $\epsilon$.
 $\mathbf{I}_M$ denotes an $M\times M$ identity matrix. 
$\mathbf{A}^*$, $\mathbf{A}^T$, $\mathbf{A}^H$, and $\mathbf{A}^\dag$ denote
complex conjugate, transpose,
conjugate
transpose, and pseudo-inverse of $\mathbf{A}$, respectively.
$[\mathbf{A}]_{m,n}$ denotes the $(m, n)^\mathrm{th}$ element of $\mathbf{A}$.
 $[\mathbf{A}]_{K\times P}$ represents $\mathbf{A}$ with $K$ rows and
$P$ columns.
 $\otimes$ and $\circ$ represent Kronecker product and Hadamard
product, respectively
and
  $\parallel\mathbf{x}\parallel_{p}$ denotes $l_p$-norm of $\mathbf{x}$.
 $diag[\mathbf{x}]$ represents a diagonal matrix having the elements
of $\mathbf{x}$ as
diagonal elements and 
$diag[\mathbf{A}]$ denotes a column vector with diagonal elements of 
$\mathbf{A}$ as its elements.
$\mathbf{Tr(A)}$ represents sum of the diagonal elements of $\mathbf{A}$ and
$\mathrm{ppm}$ represents parts per million.}
\par
So far, most studies on OFDM systems have considered synchronization
impairments and channel
separately
~\cite{morelli_tutorial , morelli2,Nugen, joint_FT_CFO, joint_FT_CFO1, Saemi, Gault}.
 Maximum Likelihood (ML)
estimation algorithms for joint estimation of CFO, channel, and STE
 of an Orthogonal Frequency Division Multiplexing Access (OFDMA) system and
Single Input Single Output (SISO)-OFDM system are
proposed in~\cite{morelli2} and~\cite{joint_FT_CFO}, respectively, but SFO is
assumed to be zero. Similarly, a joint ML time frequency
synchronization and channel estimation algorithm for MIMO-OFDM systems has been 
proposed in~\cite{joint_FT_CFO1} and~\cite{Saemi}, without considering the
effect of SFO.  
In~\cite{Nugen}, a pilot-aided joint Channel Impulse Response (CIR), CFO, and
SFO
estimation scheme has been proposed for MIMO-OFDM systems, assuming a
perfect start of
frame detection.
An ML estimator of SFO and CFO for an OFDM system is developed in
\cite{KimML}, but STE is assumed to be zero with perfect channel
knowledge. Joint ML estimators
for SFO and channel for OFDM systems are proposed in \cite{Gault}, without
considering the effect of CFO and STE.  
\par
 In this paper, we propose ML algorithms for the joint estimation of SFO, CFO,
STE, and channel in MIMO-OFDM system.
 An ML algorithm is proposed, where the multi-dimensional
optimization problem for estimating the parameters is reduced to a
two-dimensional and a one-dimensional grid search.
To reduce the complexity further, a Modified ML (MML) algorithm, 
which involves only multiple one-dimensional searches, is also proposed.
 Further, a Stage-wise ML (SML) algorithm using existing algorithms,
which estimate fewer number of parameters, is also proposed.
 Cram\'{e}r-Rao Lower Bounds (CRLB) for the estimation of continuous parameters
are derived,
which show the coupling effect among different impairments and the significance
of the joint estimation algorithms. Some results of this paper are also
presented in~\cite{renu}.
\section{SYSTEM MODEL}\label{system_model}
Consider a MIMO-OFDM system with $N_T$ transmit antennas and $N_R$ receive
antennas using Quaternary Phase Shift Keying (QPSK) Modulation and $N$
subcarriers per antenna. 
The input bit stream is first multiplexed in space and time before being grouped
by the serial-to-parallel converter. 
After Inverse Fast Fourier Transform (IFFT) operation, cyclic prefix (CP)
insertion and digital-to-analog conversion, the transmitted signal from
${u^\mathrm{th}}$ radio frequency (RF) transmit antenna undergoes fading by the
channel before reaching the ${v}^\mathrm{th}$ RF receive antenna. Let $T$ be the
sampling time of the sampling frequency oscillator at
the transmitter. Then the subcarrier spacing is
given by $1/(NT)$. The CIR between ${u^\mathrm{th}}$ transmit and
${v^\mathrm{th}}$ receive antenna is
$h_{u,v}(\tau)=\sum_{l=0}^{L_{u,v}-1}{h_{u,v,l}\delta(\tau-\tau_{l})}$, where
$h_{u,v,l}$ denotes the channel coefficient, $\tau_{l}$ denotes the
$l^\mathrm{th}$ channel path delay ($\tau_{l}=lT$), and 
$L_{u,v}$ denotes the length of CIR, for
$u=1,2,3,\hdots,N_T$ and $v=1,2,3,\hdots,N_R$. 
\par
 At each receive antenna, a superposition of faded signals from all transmit
antennas together with noise is received.  
Frequency differences between RF oscillators used in the MIMO-OFDM transmitter
and receiver, and channel induced Doppler shifts
cause a net CFO of $\Delta f_c $ in the received signal, where $f_c$ is the
operating radio carrier frequency of the RF oscillator.
 Furthermore, at the receiver, the received signal is sampled
at $T^{'}$ where $T^{'} \neq T$ and $\Delta T=T^{'}-T$, which results in SFO.
 The normalized CFO is
$\epsilon=\Delta f_cNT$
and the normalized SFO is $\eta=\Delta T/T$.
The frame arrival detection in OFDM based systems is done using different
correlation
methods
~\cite{startofframe1, startofframe2, startofframe3}. The
main drawback
of these methods is that the correlation functions do not produce a sharp peak
at the arrival of the frame, which results in difficulty to find the fine frame
time arrival instant, causing STE~\cite{joint_FT_CFO1}. Let the
STE, after start of frame detection, be given by integer number of
samples $\theta T$, where $\theta$ is the normalized STE.
The fractional part of the timing error is incorporated into the CIR 
~\cite{morelli3, timing_fraction}. 
We assume all receive antennas experience common synchronization impairments in
a single user MIMO-OFDM
system~\cite{joint_FT_CFO1, Saemi}.
\par
We consider the estimation of impairments
during the training
block. As the training blocks are usually preceded by long CP in practical
applications, we assume that the length of CP is greater than
$(L_m+\theta_{max}$), where $L_m=\max\limits_{u,v}\{L_{u,v}\}$ and
$\theta_{max}$ is the maximum STE~\cite{morelli2}. 
The received signal at
the ${v}^\mathrm{th}$ RF receive antenna is given by
$$
r_v(n)=\exp(j2\pi\Delta
f_cnT^{'})\sum_{u=1}^{N_{T}}\sum_{l=0}^{L_m-1}h_{u,v,l}s_u(nT^{'}-\theta
T-\tau_{l})+w_v(n).
$$ 
where
$s_{u}(n)$ is the signal transmitted by the $u^\mathrm{th}$ transmit antenna and
$w_v(n)$  is the complex additive Gaussian noise at the $v^\mathrm{th}$ receive
antenna with mean zero and variance $\sigma_{w}^2$.
We have,
$$ s_u(nT^{'}-\theta T-\tau_{l})=s_{u}(n(T+\Delta T)-\theta
T-\tau_{l})=s_u((n_{\eta}-\theta-l)T)$$
$$\exp(j2\pi\Delta f_cnT^{'})=\exp(j2\pi\varepsilon n(T+\Delta
T)/NT)=\exp(j2\pi\varepsilon_{\eta}n/N).$$
where $\varepsilon_\eta=\varepsilon(1+\eta)$, and
$n_\eta=n(1+\eta).$ Thus,
\begin{equation}\label{basebandsignal}
 r_v(n)=\exp(j2\pi\varepsilon_{\eta}n/N)\sum_{u=1}^{N_T}\sum_{l=0}^{L_m-1}h_{u,v
,l}s_u(n_{\eta}-\theta-l)+ w_v(n).
\end{equation} 
\par
From (\ref{basebandsignal}), it can be observed that there is coupling between
the parameters $\epsilon$, $\eta$, $\theta$ and $h_{u,v}$.   
 The Channel Frequency Response (CFR) is expressed as 
$\tilde{h}_{u,v}(k)=\sum_{l=0}^{L_{u,v}-1} h_{u,v,l}\exp(-j2\pi kl/N).$
Let $x_{u}(n)=s_{u}(n_\eta)$. The frequency domain signal samples are 
$\tilde{x}_{u}(k)=\sum_{n=0}^{N-1}x_{u}(n)\exp (-j2\pi kn/N).$  
After removal of CP, $ r_{v}(n)$ in (\ref{basebandsignal}) can be expressed as,
\begin{equation}
\label{r_v}
 r_{v}(n)=\frac{\exp(j2\pi\varepsilon_{\eta}n/N)}{N}\sum_{u=1}^{N_T}\sum_{k=0}^{
N-1}\exp(j2\pi
n_{\eta}k/N)\exp(-j2\pi\theta k/N)\tilde{h}_{u,v}(k)\tilde{x}_{u}(k)+ w_{v}(n),
\end{equation}
where the initial offsets due to CP are assumed to be zero. Let matrices $[\mathbf{F}_1]_{N\times N}$ and $[\mathbf{F}_2]_{N\times L_m}$ be
defined
as
\begin{align}
\label{F_1}
&[\mathbf{F}_1(\eta)]_{n,k}=\frac{\exp(j2\pi k(n(1+{\eta}))/N)}{N},
\\ 
\label{F_2}
\mathrm{and}\hspace{.2cm}&[\mathbf{F}_2]_{k,l}=\exp(-j2\pi lk/N)
\end{align}
where $n, k=0,1,\hdots,N-1$ and $l=0,1,\hdots,L_m-1$.
Taking $N$ samples 
of $r_{v}(n)$ in (\ref{r_v}),
\begin{align}
\label{datamodel0}
&\mathbf{r}_v=\mathbf{D}(\varepsilon,\eta)\mathbf{F}_1({\eta})\mathbf{G}
(\theta)\mathbf{X}\tilde{\mathbf{h}}_v + \mathbf{w}_v, 
\\
\label{D}
\mathrm{where}\hspace{.2cm}&\mathbf{D}(\varepsilon,\eta)=
diag[1,\exp(j2\pi\varepsilon_{\eta}/N),\hdots,\exp(j2\pi\varepsilon_{\eta}
(N-1)/N)],\\
\label{G}
&\mathbf{G}(\theta)=
diag[1,\exp(-j2\pi\theta/N),\hdots,\exp(-j2\pi(N-1)\theta/N)],
\end{align}
$[\tilde{\mathbf{h}}_{v}]_{NN_{T}\times1}=[\tilde{\mathbf{h}}^{T}_{1,v},\tilde{
\mathbf{h}}^{T}_{2,v},\hdots,\tilde{\mathbf{h}}^{T}_{N_{T},v}]^T,\hspace{.2cm}
\mathrm{and} \hspace{.2cm}[\mathbf{X}]_{N\times NN_{T}}=[\mathbf{X}_{1},
\mathbf{X}_{2},\hdots,\mathbf{X}_{N_{T}}]\hspace{.2cm} \mathrm{with}$

$\mathbf{X}_{u}=diag[\tilde{x}_u(0),
\tilde{x}_u(1),\hdots,\tilde{x}_u(N-1)],\hspace{.2cm} \mathrm{and} \hspace{.2cm}
\tilde{\mathbf{h}}_{u,v}=[\tilde{h}_{u,v}(0),
\tilde{h}_{u,v}(1),\hdots,\tilde{h}_{u,v}(N-1)]^T.$ 
We have CIR, denoted by $[\mathbf{h}_{u,v}]_{L_{m}\times1}$, and CFR, denoted by
$\tilde{[\mathbf{h}}_{u,v}]_{N\times1}$, related as
$\tilde{\mathbf{h}}_{u,v}=\mathbf{F}_2\mathbf{h}_{u,v}.$
\begin{equation}\label{datamodel0}
\mathrm{Thus,}\hspace{.2cm}
\mathbf{r}_v=\mathbf{D}(\varepsilon,\eta)\mathbf{F}_1({\eta})\mathbf{G}
(\theta)\mathbf{X}(\mathbf{I}_{N_T}\otimes \mathbf{F}_2)\mathbf{{h}}_v +
\mathbf{w}_v
\end{equation}
where $[\mathbf{h}_{v}]_{N_{T}L_{m}\times1}=[\mathbf{h}^{T}_{1,v}, 
\mathbf{h}^{T}_{2,v},\hdots,\mathbf{h}^{T}_{N_{T},v}]^T$.
Stacking the outputs of $N_R$ receive antennas, denoted by
$[\mathbf{r}]_{NN_{R}\times1}=[\mathbf{r}_{1}^T, 
\mathbf{r}_{2}^T,\hdots,\mathbf{r}_{N_R}^T ]^T$, and simplifying, 
\begin{equation}\label{datamodel}
\mathbf{r}=\mathbf{A}(\epsilon,\eta,\theta)\mathbf{h}+\mathbf{w}
\end{equation}
\begin{equation}\label{A}
\mathrm{where}
\hspace{.2cm}\mathbf{A}(\epsilon,\eta,\theta)=\mathbf{I}_{N_R}\otimes
(\mathbf{D}(\epsilon,\eta)\mathbf{F}_1(\eta)\mathbf{G}(\theta)\mathbf{X
(\mathbf{ I } } _ { N_T } \otimes
\mathbf{F}_2))
\end{equation}
and 
$[\mathbf{h}]_{N_{R}N_TL_m\times1}=[\mathbf{h}_{1}^{T},
\mathbf{h}_{2}^{T},\hdots ,
\mathbf{h}_{N_R}^T]^T.$
\par
Neglecting the effect of SFO in the system model in (\ref{datamodel}), i.e.
letting $\eta=0$, will result in the system model
as given
in
~\cite{joint_FT_CFO1} and~\cite{Saemi}, where
the effects
of CFO, STE, and channel are considered. 
Similarly, if the STE is neglected in (\ref{datamodel}), i.e. making
$\theta=0$, we obtain the system model as in
\cite{Nugen},
where the effects of CFO, SFO, and channel are considered. Also, if the effect of CFO is
not considered in (\ref{datamodel}), i.e. making $\epsilon=0$, will result in
the
 system model as given in \cite{Gault}, for a SISO-OFDM system. Thus, the
system model in
(\ref{datamodel}) is general and considers the synchronization impairments and
channel together.
\section{CRLB Analysis}\label{CRLB}
 A closed form expression of the CRLB for the estimation of CFO and channel is
derived in~\cite{stoica} and~\cite{morelli2} for the cases of single carrier
and
multicarrier communication systems, respectively. 
The related results can also be found in~\cite{morelli3} and~\cite{morelli4}.
Also, a joint estimation algorithm is described in~\cite{NugenRLS},
with CRLB derivation for the joint estimation of CFO, SFO, and channel for a SISO-OFDM system.
 But, calculation errors of equations ($28$) and ($29$) in~\cite{NugenRLS} were
reported and  re-derivation of CRLB for the joint estimation of CFO
and SFO
for a SISO-OFDM system is done in~\cite{KimML}, without considering the channel
as a parameter to be estimated.
 In this section, we obtain the CRLB for the estimation of CFO ($\epsilon$) and
SFO ($\eta$) for a MIMO-OFDM system, considering the effect of STE ($\theta$),
and analyzing the cases where channel is considered as a
parameter to be estimated.
 The parameter vector of interest is represented as
$\boldsymbol{\alpha}=[\epsilon, \eta, \mathbf{h}_R^T,\mathbf{h}_I^T]^T$,
where
$\mathbf{h}_R$ and $\mathbf{h}_I$ represent real and imaginary parts of
$\mathbf{h}$, respectively, and $\theta$ being a discrete parameter is omitted
from the parameter vector of interest. Let
$\mathbf{X}_1=\mathbf{X}(\mathbf{I}_{N_T}\otimes
\mathbf{F}_2)$, and $\boldsymbol{\mu}$ be the mean of the received signal
vector
$\mathbf{r}$ in (\ref{datamodel}). 
Then, the Fisher Information Matrix
$(\mathbf{\Gamma})$ is given by~\cite{stoica},~\cite{Kay}, 
\begin{equation}\label{FIM}
\mathbf{\Gamma}=\frac{2}{\sigma_{w}^2}\Re\left[\frac{\partial
\boldsymbol{\mu}^H}{\partial\boldsymbol{ \alpha
}}\frac{\partial\boldsymbol{\mu}}{ \partial\boldsymbol{ \alpha }^T} \right ],
\end{equation}
\begin{equation}\label{mu}
 \mathrm{where}\hspace{.5cm}\boldsymbol{ \mu }=(\mathbf{I}_{N_R}\otimes
(\mathbf{DF}_1\mathbf{GX}_1))\mathbf{h}.
\end{equation}
\subsection{CRLB without channel being a parameter to be estimated}
Let $\mathbf{\Gamma}{\hspace{-.1cm}}_{woc}$ denote $\mathbf{\Gamma}$ without
considering the channel as a parameter to be estimated.
From (\ref{FIM}), 
\begin{equation}\label{FIMwoc}
 \mathbf{\Gamma}{\hspace{-.1cm}}_{woc}=\frac{2}{\sigma_{w}^2}\Re\left[
  \begin{array}{cc}
    \gamma_{\epsilon,\epsilon} 
&\gamma_{\epsilon,\eta}  \\
   \gamma_{\eta,\epsilon}  
&\gamma_{\eta,\eta}  \\
    \end{array}
\right].
\end{equation}
\begin{equation}\label{C1}
 \mathrm{Let}\hspace{.2cm}[\mathbf{C}_1]_{N\times N}= diag[0,1,2,\hdots,(N-1)].
\end{equation}
Then from (\ref{D}) and (\ref{mu}) we have 
\begin{equation}
 \frac{\partial\mathbf{D}}{\partial\epsilon}=\frac{j2\pi}{N}(1+\eta)\mathbf{DC}
_1
\end{equation}
\begin{equation}
 \mathrm{and}\hspace{0.5cm}\label{domudoepsilon}
\frac{\partial\boldsymbol{ \mu
}}{\partial\epsilon}=\left(\mathbf{I}\otimes\left(\frac{
j2\pi(1+\eta)
}{N}\mathbf{DC}_1\mathbf{F}_1\mathbf{G}\mathbf{X}_1\right)\right)\mathbf{h}.
%
\end{equation}
Substituting (\ref{domudoepsilon}) 
in $\gamma_{\epsilon,\epsilon}=\displaystyle
\frac{\partial\boldsymbol{ \mu
}^H}{\partial\epsilon}\frac{\partial\boldsymbol{ \mu }}{\partial\epsilon}$ and
simplifying using the properties of Kronecker product and matrix
derivatives~\cite{Horn_Johnson} we get,
\begin{equation}\label{FIMepseps}
 \gamma_{\epsilon,\epsilon}=\mathbf{h}^H\left(\mathbf{I
}\otimes\left(\left(\frac{2\pi(1+\eta)}{N}\right)^2
\mathbf{X}_1^H\mathbf{G}^H\mathbf{F}_1^H\mathbf{C}_1^2\mathbf{F}_1\mathbf{GX}_1
\right)\right)\mathbf{h}.
\end{equation}
$$\mathrm{Let}\hspace{.2cm}[\mathbf{C}_2]_{N\times N}=\left(diag(\mathbf{I}_N)\otimes
[diag(\mathbf{C}_1)]^T\right)\circ\left([diag(\mathbf{I}_N)]^T\otimes
diag(\mathbf{C}_1)\right).$$ 
Then from (\ref{F_1}), (\ref{D}), and
(\ref{mu}),
\begin{equation}\label{doMudoeta}
 \frac{\partial \boldsymbol{ \mu
}}{\partial\eta}=\left(\mathbf{I}\otimes\frac{\partial
(\mathbf{DF}_1\mathbf{GX}_1)}{\partial\eta}\right)\mathbf{h}=\left(\mathbf{I}
\otimes\left(\frac{\partial
\mathbf{D}}{\partial\eta}\mathbf{F}_1\mathbf{GX}_1+\mathbf{D}\frac{
\partial\mathbf { F}_1 } {
\partial\eta}\mathbf{GX}_1\right)\right)\mathbf{h}
\end{equation}
$$
\mathrm{where}\hspace{.2cm}\frac{\partial\mathbf{D}}{\partial\eta}=\frac{
j2\pi\epsilon}{N}\mathbf{DC}_1,\hspace{.2cm}
\mathrm{and}\hspace{.2cm}\frac{\partial\mathbf{F}_1}{\partial\eta}=\frac{j2\pi}
{
N}(\mathbf{C}_2\circ\mathbf{F}_1).
$$
Using (\ref{domudoepsilon}) and (\ref{doMudoeta}), we obtain the closed form
expressions for 
$\gamma_{\epsilon,\eta}=\displaystyle
\frac{\partial\boldsymbol{ \mu
}^H}{\partial\epsilon}\frac{\partial\boldsymbol{ \mu }}{\partial\eta}$, 
$\gamma_{\eta,\epsilon}=\displaystyle
\frac{\partial\boldsymbol{ \mu
}^H}{\partial\eta}\frac{\partial\boldsymbol{ \mu }}{\partial\epsilon}$ and
$\gamma_{\eta,\eta}=\displaystyle
\frac{\partial\boldsymbol{ \mu
}^H}{\partial\eta}\frac{\partial\boldsymbol{ \mu }}{\partial\eta}$ as,  
\begin{align}\label{FIMepseta}
 \gamma_{\epsilon,\eta}=\gamma_{\eta,\epsilon}^H=\mathbf{h}^H\left(\mathbf{I
}\otimes\left(\left(\frac{
4\pi^2(1+\eta)}{N^2}\right)
\mathbf{X}_1^H\mathbf{G}^H\mathbf{F}_1^H\mathbf{C}_1(\epsilon
\mathbf{C}_1\mathbf{F}_1
+(\mathbf{C}_2\circ\mathbf{F}_1)\big)\mathbf{
GX}_1 \right)\right)\mathbf{h}
\end{align}
\begin{align}\label{FIMetaeta}
 \gamma_{\eta,\eta}=\displaystyle
\mathbf{h}^H\Bigg(\mathbf{I}\otimes\Big(\left(\frac{2\pi}{N}\right)^2\mathbf{X
} ^H_1\mathbf{G}^H
\Big(\epsilon\mathbf{F}^H_1\mathbf{C}
_1(\mathbf{C}_2\circ\mathbf{F}_1)
&+(\mathbf{C}_2\circ\mathbf{F}_1)^H(\mathbf{C}
_2\circ\mathbf{F}_1)
+\epsilon^2\mathbf { F } ^H_1\mathbf{C}^2
_1\mathbf{F}_1\nonumber\\
&+\epsilon(\mathbf{C}_2\circ\mathbf{F}_1)^H\mathbf{C}
_1\mathbf{F}_1\Big)\mathbf{GX}_1\Big)\Bigg)
\mathbf{h}
\end{align}
\par
Using (\ref{FIMepseps}), (\ref{FIMepseta}), and (\ref{FIMetaeta}), the
CRLBs for the estimation of $\epsilon$ and $\eta$, without channel being
considered as a parameter to be estimated, denoted by
$\mathrm{CRLB}(\epsilon_{woc})$ and
$\mathrm{CRLB}(\eta_{woc})$, respectively, are obtained as,
\begin{align}\label{epswoc}
 &\mathrm{CRLB}(\epsilon_{woc})=\frac{\gamma_{\eta,\eta}}{\gamma_{\epsilon,
\epsilon }
\gamma_{\eta,\eta}-\gamma_{\eta,\epsilon}
\gamma_{\epsilon,\eta}}\\
\label{etawoc}
 \mathrm{and}\hspace{0.5cm}&\mathrm{CRLB}(\eta_{woc})=\frac{\gamma_{\epsilon,
\epsilon}}{
\gamma_ { \epsilon,\epsilon}
\gamma_{\eta,\eta}-\gamma_{\eta,\epsilon}
\gamma_{\epsilon,\eta}}.
\end{align}
\subsection{CRLB with channel being a parameter to be estimated}
Let $\mathbf{\Gamma}{\hspace{-.1cm}}_{wc}$ denote $\mathbf{\Gamma}$ considering
the channel as a parameter to be estimated.
 From (\ref{FIM}) and (\ref{FIMwoc}), 

\begin{equation}\label{FIMwc}
\mathbf{\Gamma}{\hspace{-.1cm}}_{wc}=\frac{2}{\sigma_{w}^2}\Re\begin{bmatrix}
\mathbf{\Gamma}{\hspace{-.1cm}}_{woc}& 
\begin{array}{cc}
\boldsymbol{\gamma}_{\epsilon,\mathbf{h}_R}
&\boldsymbol{\gamma}_{\epsilon,\mathbf{h}_I} \\
\boldsymbol{\gamma}_{\eta,\mathbf{h}_R}
&\boldsymbol{\gamma}_{\eta,\mathbf{h}_I}
\end{array}
\\
\begin{array}{ccc}
\boldsymbol{\gamma}_{\mathbf{h}_R,\epsilon}&
\boldsymbol{\gamma}_{\mathbf{h}_R,\eta} \\
\boldsymbol{\gamma}_{\mathbf{h}_I,\epsilon} &
\boldsymbol{\gamma}_{\mathbf{h}_I,\eta}
\end{array}
& \mathbf{\Gamma}{\hspace{-.1cm}}_{\mathbf{h},\mathbf{h}} 
\end{bmatrix},\\ 
\end{equation}
\begin{equation}
 \mathrm{where}\hspace{.2cm}\mathbf{\Gamma}{\hspace{-.1cm}}_{\mathbf{h,h}}=\left
[
  \begin{array}{cc}
    \mathbf{\Gamma}{\hspace{-.1cm}}_{\mathbf{h}_R,\mathbf{h}_R} 
&\mathbf{\Gamma}{\hspace{-.1cm}}_{\mathbf{h}_R,\mathbf{h}_I} \\
   \mathbf{\Gamma}{\hspace{-.1cm}}_{\mathbf{h}_I,\mathbf{h}_R} 
&\mathbf{\Gamma}{\hspace{-.1cm}}_{\mathbf{h}_I,\mathbf{h}_I} \\
      \end{array}
\right].
\end{equation}
From (\ref{mu}) we have,
\begin{equation}\label{domudohr}
 \frac{\partial\boldsymbol{ \mu }}{\partial\mathbf{h}_R}=\mathbf{I}\otimes
(\mathbf{DF}_1\mathbf{GX}_1), 
\end{equation}
\begin{equation}\label{domumdohi}
 \mathrm{and}\hspace{0.5cm}\frac{\partial\boldsymbol{ \mu
}}{\partial\mathbf{h}_I}=j(\mathbf{I}
\otimes (\mathbf{DF}_1\mathbf{GX}_1))
=j\frac{\partial\boldsymbol{ \mu }}{\partial\mathbf{h}_R} 
\end{equation}
Substituting (\ref{domudoepsilon}), (\ref{domudohr}), and (\ref{domumdohi}) in 
$\mathbf{\Gamma}{\hspace{-.1cm}}_{\mathbf{h}_R,\mathbf{h}_R}=\displaystyle
\frac{\partial\boldsymbol{ \mu
}^H}{\partial\mathbf{h}_R}\frac{\partial\boldsymbol{ \mu }}{\partial\mathbf{h}
_R^T}$,
$\mathbf{\Gamma}{\hspace{-.1cm}}_{\mathbf{h}_I,\mathbf{h}_I}=\displaystyle\frac{
\partial\boldsymbol{ \mu
}^H}{\partial\mathbf{h}_I}\frac{\partial\boldsymbol{ \mu
}}{\partial\mathbf{h}_I^T}$,
$\mathbf{\Gamma}{\hspace{-.1cm}}_{\mathbf{h}_R,\mathbf{h}_I}=\displaystyle\frac{
\partial\boldsymbol{ \mu
}^H}{\partial\mathbf{h}_R}\frac{\partial\boldsymbol{ \mu
}}{\partial\mathbf{h}_I^T}$, 
$\mathbf{\Gamma}{\hspace{-.1cm}}_{\mathbf{h}_I,\mathbf{h}_R}=\displaystyle\frac{
\partial\boldsymbol{ \mu
}^H}{\partial\mathbf{h}_I}\frac{\partial\boldsymbol{ \mu
}}{\partial\mathbf{h}_R^T}$, 
$\mathbf{\Gamma}{\hspace{-.1cm}}_{\epsilon,\mathbf{h}_R}=\displaystyle
\frac{\partial\boldsymbol{ \mu
}^H}{\partial\epsilon}\frac{\partial\boldsymbol{ \mu
}}{\partial\mathbf{h}_R^T}$,
$\mathbf{\Gamma}{\hspace{-.1cm}}_{\mathbf{h}_R,\epsilon}=\displaystyle\frac{
\partial\boldsymbol{ \mu
}^H}{\partial\mathbf{h}_R}\frac{\partial\boldsymbol{ \mu }}{\partial\epsilon}$,
$\mathbf{\Gamma}{\hspace{-.1cm}}_{\epsilon,\mathbf{h}_I}=\displaystyle\frac{
\partial\boldsymbol{ \mu
}^H}{\partial\epsilon}\frac{\partial\boldsymbol{ \mu
}}{\partial\mathbf{h}_I^T}$,
and 
$\mathbf{\Gamma}{\hspace{-.1cm}}_{\mathbf{h}_I,\epsilon}=\displaystyle\frac{
\partial\boldsymbol{ \mu
}^H}{\partial\mathbf{h}_I}\frac{\partial\boldsymbol{ \mu }}{\partial\epsilon}$,
and simplifying we get,
\begin{align}
\label{FIMhrhr}
 &\mathbf{\Gamma}{\hspace{-.1cm}}_{\mathbf{h}_R,\mathbf{h}_R}=\mathbf{I}
\otimes\left(\mathbf{X}_1^H\mathbf{G}^H\mathbf{F}_1^H\mathbf{F}_1\mathbf{GX}_1
\right)=\mathbf{
\Gamma}{\hspace{-.1cm}}_{\mathbf{h}_I,\mathbf{h}_I}=-j\mathbf{\Gamma}{\hspace{
-.1cm}}_{\mathbf{h}_R,\mathbf{h}_I}=j\mathbf{\Gamma}{\hspace{-.1cm}}_{\mathbf{h}
_I,\mathbf{h}_R},\\
\label{FIMepshr}
&\mathbf{\Gamma}{\hspace{-.1cm}}_{\epsilon,\mathbf{h}_R}=\mathbf{h}
^H\left(\mathbf{I}\otimes\left(\frac{(-j2\pi(1+\eta))}{N}\mathbf{X}_1^H\mathbf{
G}
^H\mathbf{F}_1^H\mathbf{C}_1\mathbf{F}_1\mathbf{GX}_1\right)\right)=\mathbf{
\Gamma } { \hspace { -.05cm
} } _ {
\mathbf{h}_R,\epsilon}^H=\frac{\mathbf{\Gamma}{\hspace{-.1cm}}_{\epsilon,
\mathbf { h }
_I}}{j}=\frac{\mathbf{\Gamma}{\hspace{-.05cm}}_{\mathbf{h}_I,\epsilon}^H}{j}.
\end{align}
Similarly, using (\ref{doMudoeta}), (\ref{domudohr}), and (\ref{domumdohi}) we
obtain the closed form expressions for 
$\mathbf{\Gamma}{\hspace{-.1cm}}_{\eta,\mathbf{h}_R}=\displaystyle
\frac{\partial\boldsymbol{ \mu
}^H}{\partial\eta}\frac{\partial\boldsymbol{ \mu }}{\partial\mathbf{h}_R}$,
$\mathbf{\Gamma}{\hspace{-.1cm}}_{\mathbf{h}_R,\eta}=\displaystyle
\frac{\partial\boldsymbol{ \mu
}^H}{\partial\mathbf{h}_R}\frac{\partial\boldsymbol{ \mu }}{\partial\eta}$,
$\mathbf{\Gamma}{\hspace{-.1cm}}_{\eta,\mathbf{h}_I}=\displaystyle
\frac{\partial\boldsymbol{ \mu
}^H}{\partial\eta}\frac{\partial\boldsymbol{ \mu }}{\partial\mathbf{h}_I}$,
and
$\mathbf{\Gamma}{\hspace{-.1cm}}_{\mathbf{h}_I,\eta}=\displaystyle
\frac{\partial\boldsymbol{ \mu
}^H}{\partial\mathbf{h}_I}\frac{\partial\boldsymbol{ \mu }}{\partial\eta}$ as,
\begin{align}
 \label{FIMetahr}
\mathbf{\Gamma}{\hspace{-.1cm}}_{\mathbf{h}_R,\eta}&=\left(\mathbf{I}
\otimes\left(\left(\frac{j
2\pi}{N}\right)
\mathbf{X}^H_1\mathbf{G}^H\mathbf{F}^H_1\left(\epsilon\mathbf{C}_1
\mathbf{F}_1+\left(\mathbf{C}_2\circ\mathbf{F}_1\right)\right)\mathbf{GX}
_1 \right)\right)\mathbf{h}
=\frac{{\mathbf{\Gamma}{\hspace{
-.05cm}}_{\eta,\mathbf{h}_I}^H}}{-j}=\mathbf{\Gamma}{\hspace{-.05cm}}_{
\eta,\mathbf{h}_R }
^H=\frac{{\mathbf{\Gamma}{\hspace{-.05cm}}_{\mathbf{h}_I,\eta}}}{-j}. 
\end{align}
\par
Using (\ref{FIMwoc}), (\ref{FIMhrhr}), (\ref{FIMepshr}), and (\ref{FIMetahr}),
the CRLBs for the estimation of $\epsilon$ and $\eta$, with channel being
considered a parameter to be estimated, denoted by
$\mathrm{CRLB}(\epsilon_{wc})$ and
$\mathrm{CRLB}(\eta_{wc})$, respectively, are obtained as,
\begin{align}\label{epswc}
 &\mathrm{CRLB}(\epsilon_{wc})=[\mathbf{\Gamma}{\hspace{-.009cm}}_{wc}^{-1}]_{1,
1},\\
\label{etawc}  
 \mathrm{and}\hspace{0.5cm}&\mathrm{CRLB}(\eta_{wc})=[\mathbf{\Gamma}{\hspace{
-.009cm}}_{
wc}^{-1}]_{2,2}.
\end{align}
Also, using MATLAB notation, we have CRLB for the estimation of $\mathbf{h}$ as,
  \begin{equation}\label{hwc}
\mathrm{CRLB}(\mathbf{h})=[\mathbf{\Gamma}{\hspace{-.009cm}}_{wc}^{-1}]_{
3:(2L_m+2),
3:(2L_m+2)}.
  \end{equation}
   \subsection{Significance of Joint Estimation}
\begin{figure}[htb]
\centering{
\resizebox{9.5cm}{7.5cm}{
\includegraphics[]{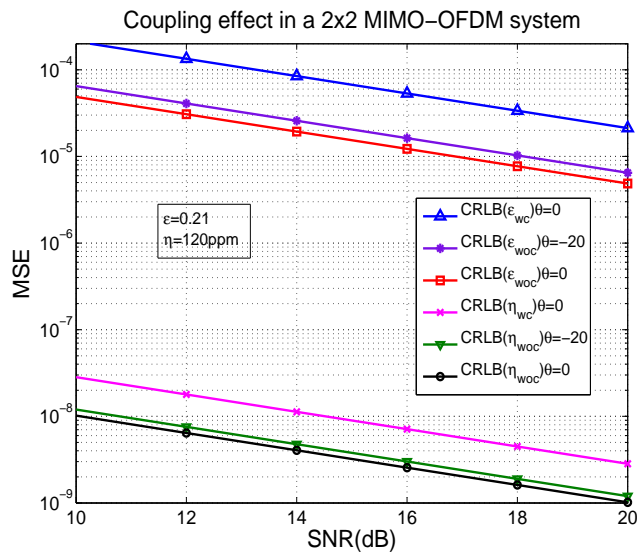}}}
\caption{CRLB for the estimation of $\epsilon$ and $\eta$ as a function of SNR(dB)
showing the coupling effect.}%
\label{coupling}
\end{figure}

The non-zero off-diagonal elements in (\ref{FIMwc}) show
the coupling between the impairments, CFO, SFO, STE, and channel, as observed in
(\ref{basebandsignal}), which emphasizes the
significance of a joint estimator.  A study of coupling between CRLBs for the
estimation of CFO and channel is explained in~\cite{stoica}, without
considering the effect of SFO and STE.
 In this section, we present coupling effects due to STE and
channel, in the CRLBs for the estimation of $\epsilon$ and $\eta$
as shown in Fig.\ref{coupling}. 
The CRLB equations in (\ref{epswoc}), (\ref{etawoc}), (\ref{epswc}), and
(\ref{etawc}) are evaluated for a $2\times2$ MIMO-OFDM system with $N=128$, and 
 $L_m=6$,
having impairment values, $\epsilon=0.21$, $\eta=120$ $\mathrm{ppm}$, and
$\theta=0$ or $-20$,
as shown in the figure.
The derived CRLB expressions depend on the specific channel
realization. Therefore, the CRLB expressions are numerically averaged over
$10^3$ channel realizations, using Spatial Channel Model (SCM) specified by the
${3^\mathrm{rd}}$ Generation
Partnership Project (3GPP)~\cite{SCM1}.
For the above given simulation setup, the change in
CRLBs plotted in Fig.\ref{coupling} can be approximately represented as,
\begin{align}
 &\mathrm{SNR}\hspace{.1cm}\mathrm{at}\hspace{.1cm}\mathrm{CRLB}(\epsilon_{wc}
)|_{
\theta=0}\approx\mathrm{SNR}\hspace{.1cm}\mathrm{at}\hspace{.1cm}\mathrm{CRLB}
(\epsilon_
{woc}
)|_{
\theta=0}+6.25\hspace{.1cm}\mathrm{ dB } \nonumber\\
&\mathrm{SNR}\hspace{.1cm}\mathrm{at}\hspace{.1cm}\mathrm{CRLB}(\eta_{wc}
)|_{
\theta=0}\approx\mathrm{SNR}\hspace{.1cm}\mathrm{at}\hspace{.1cm}\mathrm{CRLB}
(\eta_{woc
}
)|_{
\theta=0}+4.5\hspace{.1cm}\mathrm{ dB } \nonumber\\
&\mathrm{SNR}\hspace{.1cm}\mathrm{at}\hspace{.1cm}\mathrm{CRLB}(\epsilon_{woc}
)|_{
\theta=-20}\approx\mathrm{SNR}\hspace{.1cm}\mathrm{at}\hspace{.1cm}\mathrm{CRLB}
(\epsilon_{woc}
)|_{\theta=0} +1.25\hspace{.1cm}\mathrm{ dB } \nonumber\\
&\mathrm{SNR}\hspace{.1cm}\mathrm{at}\hspace{.1cm}\mathrm{CRLB}(\eta_{woc})|_{
\theta=-20}
\approx\mathrm{SNR}\hspace{.1cm}\mathrm{at}\hspace{.1cm}\mathrm{CRLB}(\eta_{woc}
)|_{\theta=0}
+1\hspace{.1cm}\mathrm{ dB } \nonumber
\end{align}
The above expressions show the effect of coupling of channel and STE on the
estimation of CFO and SFO, respectively. Due to the coupling effect, the
estimation of parameters without considering all impairments together results
in performance degradation. 
\section{Maximum Likelihood Estimation}\label{ML}
The ML cost function~\cite{morelli2},~\cite{Kay} of the parameters
$\epsilon,\theta,\eta,$ and $\mathbf{h}$, obtained from (\ref{datamodel})
 is, 
\begin{align}
\arg\max_{\epsilon,\eta,\theta,\mathbf{h}}P(\mathbf{r}|\epsilon,\theta,\eta,
\mathbf{h})=\arg\max_{\epsilon,\eta,\theta,\mathbf{h}}\frac{1}{(\pi\sigma_{w}
^2)^{NN_R}}\exp\left\{ \frac{-\parallel
{\mathbf{r}-\mathbf{A}\mathbf{h}}\parallel^{2}_{2}}{\sigma_{w}^{2}}\right\}
.\nonumber
\end{align}
Computing the log-likelihood function and simplifying, we obtain an equivalent
cost function,
\begin{equation}\label{costfunction}
\arg\min_{\epsilon,\eta,\theta,\mathbf{h}}J(\epsilon,\eta,\theta,\mathbf{h}
|\mathbf{r})=\arg\min_{\epsilon,\eta,\theta,\mathbf{h}}(\mathbf{r}-\mathbf{A}
\mathbf{h})^{H}(\mathbf{r}-\mathbf{A}\mathbf{h}).
\end{equation}
The multi-dimensional minimization in (\ref{costfunction}) gives the estimate of
the parameters, ${\epsilon},{\theta},{\eta}$, and $\mathbf{h}$, which is
practically not feasible. With perfect channel knowledge at receiver, the
optimization
problem
in (\ref{costfunction}) reduces to a three-dimensional minimization problem as
$\displaystyle \arg\min_{\epsilon
,\eta,\theta}J(\epsilon,\eta,\theta|\mathbf{r},\mathbf{h})
$, which also turns out to be too complex for practical purposes. In this
section, we propose an ML algorithm in which the multi-dimensional optimization
problem in (\ref{costfunction}) is reduced to a two-dimensional and a
one-dimensional grid search, by rearranging the system model.
Also, we propose a low complexity MML algorithm,
which involves multiple one-dimensional searches, by making an approximation.
Further, SML algorithm using different existing
algorithms, which estimate fewer number of parameters, is also proposed. 
\subsection{Proposed ML Algorithm}
The system model in (\ref{datamodel}) can be rewritten as 
\begin{align}\label{datamodel2}
\mathbf{r}&=\left(\mathbf{I}_{N_R}\otimes
\left(\mathbf{DF}_1\mathbf{GX(\mathbf{I}}_{N_T}\otimes
\mathbf{F}_2)\right)\right)\mathbf{{h}}+\mathbf{w}\\
\label{datamodel2a}
&=\left(\mathbf{I}_{N_R}\otimes
\left(\mathbf{DF}_1\mathbf{GX(\mathbf{I}}_{N_T}\otimes
\mathbf{F}_{2\theta_{max}})\right)\right)\mathbf{{h}}\hspace{-.08cm}
^\prime+\mathbf { w },\\
\mathrm{where}\hspace{.2cm}
\label{F_2thetamax}
&[\mathbf{F}_{2\theta_{max}}]_{k,l}=\exp(-j2\pi lk/N),
\\
\mathrm{and}\hspace{.2cm}&\mathbf{h}{\hspace{-.08cm}}^{\prime}=[\mathbf{h}{
\hspace{-.08cm}}^{\prime{ \hspace{.05cm}} T}_{\hspace{.05cm}1},
\mathbf{h}{\hspace{-.08cm}}^{\prime{
\hspace{.05cm}} T}_{\hspace{.05cm}2},\hdots,
\mathbf{h}{\hspace{-.08cm}}^{\prime{
\hspace{.05cm}} T}_{\hspace{.05cm}N_R}]^T,
\end{align}
with $\mathbf{h}{\hspace{-.08cm}}^{\prime{
\hspace{.05cm}}}_{\hspace{.05cm}v}=[\mathbf{h}{\hspace{-.08cm}}^{\prime{
\hspace{.05cm}} T}_{\hspace{.05cm}1,v}, 
\mathbf{h}{\hspace{-.08cm}}^{\prime{
\hspace{.05cm}}
T}_{\hspace{.05cm}2,v},\hdots,\mathbf{h}{\hspace{-.08cm}}^{\prime{
\hspace{.05cm}} T}_{\hspace{.05cm}N_T,v}]^T$, and 
$\mathbf{h}{\hspace{-.08cm}}^{\prime{
\hspace{.05cm}}}_{\hspace{.05cm}u,v}=[\mathbf{h}^T_{u,v} \hspace{.2cm}
\mathbf{0}^T_{\theta_{max}\times 1}]^T$
for $k=0,1,\hdots,N-1$ and $l=0,1,\hdots,(L+\theta_{max}-1)$.
The system model in (\ref{datamodel2a}) can also be represented as,
\begin{align}\label{datamodel2b}
\mathbf{r}&=\left(\mathbf{I}_{N_R}\otimes\left(
\mathbf{DF}_1\mathbf{X(\mathbf{I}}_{N_T}\otimes
\mathbf{F}_{2\theta_{max}})\right)\right)\mathbf{{h}}_\theta+\mathbf{w}\\
\label{datamodel4}
&=\mathbf{A}_1 \mathbf{{h}}_\theta +\mathbf{w},\\
\mathrm{where}\hspace{.2cm}
\label{A1}
&\mathbf{A}_1=\mathbf{I}_{N_R}\otimes
(\mathbf{DF}_1\mathbf{X(\mathbf{I}}_{N_T}\otimes
\mathbf{F}_{2\theta_{max}})),\\
\mathrm{and}\hspace{.2cm}&\mathbf{h}_{\theta}=[\mathbf{h}_{1_{\theta}}^T,
\mathbf{h}_{2_{\theta}}^T,\hdots,
\mathbf{h}_{N_{R \theta}}^T]^T,
\end{align}
with $\mathbf{h}_{v_\theta}=[\mathbf{h}_{(1,v)_\theta}^T, 
\mathbf{h}_{(2,v)_\theta}^T,\hdots,\mathbf{h}_{(N_T,v)_\theta}^T]^T$, and 
$\mathbf{h}_{(u,v)_\theta}=[\mathbf{0}^T_{\theta \times 1}
\hspace{.2cm}
\mathbf{h}_{u,v}^T \hspace{.2cm}
\mathbf{0}^T_{(\theta_{max}-\theta)\times 1}]^T$,\\
for $u=0,1,\hdots,N_T-1,$ and $v=0,1,\hdots,N_R-1$.
 From (\ref{datamodel4}), the least squares estimate of $\mathbf{{h}}_\theta$ is
given by,
$\hat{\mathbf{{h}}}_\theta={\mathbf{A}^{\dag}_1}\mathbf{r}$. Let
$\Paa$ be the projection matrix of $\mathbf{A}_1$. Therefore, the estimate of
$\epsilon$ and $\eta$ can be obtained from (\ref{datamodel4}) as,
\begin{align}
 \displaystyle
\label{2Dcostfunction}
[\hat{\epsilon}_{ML},\hat{\eta}_{ML}]&=\arg\min_{\epsilon,\eta}(\mathbf{r}
-\mathbf{A}_1 \mathbf{A}^{\dag}_1\mathbf{r})^H(\mathbf{r}
-\mathbf{A}_1 \mathbf{A}^{\dag}_1\mathbf{r})\\
\label{2Dsearch}
&=\arg\max_{\epsilon,\eta}\parallel \Paa\mathbf{r}\parallel_{2}^2\\
&=\arg\max_{\epsilon,\eta}J_1(\epsilon,\eta|\mathbf{r})\nonumber
\end{align}
\begin{equation}\label{J1}
 \mathrm{where}\hspace{.2cm}J_1(\epsilon,\eta|\mathbf{r})=\parallel
\Paa\mathbf{r}\parallel_{2}^2.
\end{equation}
From (\ref{datamodel}), the least squares estimate of $\mathbf{{h}}$ is
given by,
$\hat{\mathbf{{h}}}={\mathbf{A}^{\dag}}\mathbf{r}$. Let
$\Pa$ be the projection matrix of $\mathbf{A}$. Therefore, using the estimates
of $\epsilon$ and $\eta$ from (\ref{2Dsearch}), the estimate of
$\theta$ can be obtained as,
 \begin{align}
 \displaystyle
[\hat{\theta}_{ML}]&=\arg\min_{\theta}(\mathbf{r}
-\mathbf{A} \mathbf{A}^{\dag}\mathbf{r})^H(\mathbf{r}
-\mathbf{A} \mathbf{A}^{\dag}\mathbf{r})\\
\label{1Dsearch}
&=\arg\max_{\theta}\parallel \Pa\mathbf{r}\parallel_{2}^2\\
&=\arg\max_{\theta}J_2(\theta|\mathbf{r},\hat{\epsilon}_{ML},\hat{\eta}_{ML}
)\nonumber
\end{align}
\begin{equation}\label{J2}
 \mathrm{where}\hspace{.2cm}J_2(\theta|\mathbf{r},\hat{\epsilon}_{ML},\hat{\eta
}
_{ML})=\parallel
\Pa\mathbf{r}\parallel_{2}^2.
\end{equation}
Finally, using the estimates
of $\epsilon$, $\eta$, and $\theta$, we get the estimate of
$\mathbf{h}$ as,
\begin{align}\label{hhatML}
 [\hat{\mathbf{h}}_{ML}]=\hat{\mathbf{A}}^\dag\mathbf{r}
\end{align}
\begin{equation}\label{Ahat}
 \mathrm{where}\hspace{.2cm}\hat{\mathbf{A}}=\mathbf{I}_{N_R}\otimes
\left(\mathbf{D
} (\hat {\epsilon}_{ML
},\hat{\eta}_{ML})\mathbf{F}
_1(\hat{\eta}_{ML})\mathbf{G}(\hat{\theta}_{ML})\mathbf{X}\left(\mathbf{I}_{
N_T }
\otimes
\mathbf{F}_2\right)\right).
\end{equation}
The series of steps involved are given in Algorithm $1$.
\begin{algorithm}
\label{MLAlgo}
\caption{ML Algorithm}
{\bf Inputs:} $N$, $N_T$, $N_R$, $\mathbf{r}$, $\mathbf{X}$,
$L_m$, $[\theta_{min}, \theta_{max}, \theta_{grid}]$, $[\epsilon_{min},
\epsilon_{max},  \epsilon_{grid}]$, $[\eta_{min}, \eta_{max}, \eta_{grid}]$
\label{OMPAlgo}
 \begin{algorithmic}[1]
  \STATE Evaluate $\mathbf{F}_2$, $\mathbf{F}_{2\theta_{max}}$;\hfill $\star$
using (\ref{F_2}), (\ref{F_2thetamax})

      \FOR{$i=$ $\epsilon_{min}:\epsilon_{grid}:\epsilon_{max}$}
      \FOR{$j=$ $\eta_{min}:\eta_{grid}:\eta_{max}$}
	\STATE Evaluate $\mathbf{D}(i,j)$, $\mathbf{F}_1(j)$;\hfill $\star$
using (\ref{D}), (\ref{F_1})
\STATE $\mathbf{A}_1(i,j)=\mathbf{I}_{N_R}\otimes
\left(\mathbf{D}(i,j)\mathbf{F}_1(j)\mathbf{X}(\mathbf{I}_{N_T}\otimes
\mathbf{F}_{2\theta_{max}})\right)$
;\hfill $\star$ using (\ref{A1})
\STATE $J_1(i,j|\mathbf{r})=\parallel \Paa\mathbf{r}\parallel_{2}^2$;\hfill
$\star$ using (\ref{J1})
      \ENDFOR
    \ENDFOR
  \STATE $\displaystyle
[\hat{\epsilon}_{ML},\hat{\eta}_{ML}]=\arg\max_{i,j}J_1(i,j|\mathbf{r})$;
\STATE Evaluate $\mathbf{D}(\hat{\epsilon}_{ML},\hat{\eta}_{ML})$,
$\mathbf{F}_1(\hat{\eta}_{ML})$;\hfill $\star$
using (\ref{D}), (\ref{F_1})
\FOR{$k=$ $\theta_{min}:\theta_{grid}:\theta_{max}$}
\STATE Evaluate $\mathbf{G}(k)$,
$\mathbf{A}(\hat{\epsilon}_{ML},\hat{\eta}_{ML},k)$;\hfill $\star$ using
(\ref{G}), (\ref{A})
\STATE $J_2(k|\mathbf{r},\hat{\epsilon}_{ML
},\hat{\eta}_{ML})=\parallel \Pa\mathbf{r}\parallel_{2}^2$;\hfill $\star$ using
(\ref{J2})
\ENDFOR
\STATE $\displaystyle
[\hat{\theta}_{ML}]=\arg\max_{k}J_2(k|\mathbf{r},\hat{\epsilon}_{ML
},\hat{\eta}_{ML})$;
\STATE Evaluate $\mathbf{G}(\hat{\theta}_{ML})$;\hfill $\star$ using (\ref{G})
\STATE
$\hat{\mathbf{A}}=\mathbf{I}_{N_R}\otimes
\left(\mathbf{D
} (\hat {\epsilon}_{ML
},\hat{\eta}_{ML})\mathbf{F}
_1(\hat{\eta}_{ML})\mathbf{G}(\hat{\theta}_{ML})\mathbf{X}(\mathbf{I}_{N_T}
\otimes\mathbf{
F}_2)\right)$;\hfill $\star$ using (\ref{Ahat})
\STATE
$[\hat{\mathbf{h}}_{ML}]=\hat
{\mathbf{A}}^\dag\mathbf{r}$;\hfill $\star$ using (\ref{hhatML})
\end{algorithmic}
{\bf Output:}
$[\hat{\theta}_{ML},
\hat{\epsilon}_{ML}, \hat{\eta}_{ML}, \hat{\mathbf{h}}_{ML}]$ 
 \end{algorithm}
\subsection{Proposed Modified ML (MML) Algorithm}
\begin{algorithm}[]
\label{MLAlgo}
\caption{MML Algorithm}
{\bf Inputs:} $N$, $N_T$, $N_R$, $\mathbf{r}$, $\mathbf{X}$,
$L_m$, $[\theta_{min}, \theta_{max}, \theta_{grid}]$, $[\eta_{min}, \eta_{max},
\eta_{grid}]$
\label{OMPAlgo}
 \begin{algorithmic}[1]
    \STATE Evaluate $\mathbf{F}_2$, $\mathbf{F}_{2\theta_{max}}$,
$\mathbf{C}_1$;\hfill $\star$
using (\ref{F_2}), (\ref{F_2thetamax}), (\ref{C1})
\STATE $\mathbf{R}=diag(\mathbf{r})$;
        \FOR{$j=$ $\eta_{min}:\eta_{grid}:\eta_{max}$}
	\STATE Evaluate $\mathbf{F}_1(j)$, $\mathbf{A}_2(j)$;\hfill $\star$
using (\ref{F_1}), (\ref{A2})
\STATE
$\mathbf{C}=\mathbf{R}^H\left(\mathbf{I}_{NN_R}
-\left(\mathbf{I}_{N_R}\otimes\Pab\right)\right)$;\hfill $\star$ using (\ref{C})
\STATE
$\mathbf{c}_1(j)=\displaystyle
\frac{2\pi(1+j)}{N}diag(\mathbf{I}_{N_R}\otimes\mathbf{C}_1)$;\hfill
$\star$ using (\ref{c1})
\STATE
$\hat{\epsilon}(j)=\displaystyle
\frac{\mathbf{c}^T_1\Im(\mathbf{C}\mathbf{C}^H)\mathbf{1}}{\mathbf{c}
^T_1\Re(\mathbf{C}\mathbf{C}^H)\mathbf{c}_1}$;\hfill
$\star$ using (\ref{eps_hat})
\STATE
Evaluate $J_3(j|\mathbf{r})$;\hfill $\star$ using
(\ref{J3})
          \ENDFOR
  \STATE $\displaystyle
[\hat{\eta}_{MML}]=\arg\min_{j}J_3(j|\mathbf{r})$;\hfill
$\star$ using (\ref{etahat})
\STATE $[\hat{\epsilon}_{MML}]=\hat{\epsilon}(\hat{\eta}_{MML})$;\hfill
$\star$
evaluation similar to step 7
\STATE Evaluate $\mathbf{D}(\hat{\epsilon}_{MML},\hat{\eta}_{MML})$,
$\mathbf{F}_1(\hat{\eta}_{MML})$;\hfill $\star$
using (\ref{D}), (\ref{F_1})
\STATE Remaining steps are same as steps from $11$ to $18$ in Algorithm $1$.
\end{algorithmic}
{\bf Output:}
$[\hat{\theta}_{MML},
\hat{\epsilon}_{MML}, \hat{\eta}_{MML}, \hat{\mathbf{h}}_{MML}]$ 
 \end{algorithm}
The estimation of $\epsilon$ and $\eta$, using (\ref{2Dsearch}), by a two
dimensional grid search is not desirable for practical applications. 
 Hence, a reduced complexity MML algorithm is obtained in this
section.
From (\ref{A1}), we have 
\begin{align}
\label{A1a}
 \mathbf{A}_1&=\mathbf{I}_{N_R}\otimes (\mathbf{DF_1X(\mathbf{I}}_{N_T}\otimes
\mathbf{F}_{2\theta_{max}}))\nonumber\\
&=\mathbf{I}_{N_R}\otimes \mathbf{D}\mathbf{A}_2,\\
\label{A2}
 \mathrm{where}\hspace{0.5cm}&\mathbf{A}_2=\mathbf{F}_1\mathbf{X(\mathbf{I}}_{
N_T}\otimes
\mathbf{F}_{2\theta_{max}}).
\end{align}
 Therefore, from
(\ref{datamodel2b}), (\ref{A1}), and (\ref{A1a}), the least squares estimate of
$\mathbf{{h}}_\theta$ is
given by,
\begin{align}
 \hat{\mathbf{{h}}}_\theta&={\mathbf{A}^{\dag}_1}\mathbf{r}\nonumber\\
&=(\mathbf{A}^{H}_1\mathbf{A}_1)^{-1}\mathbf{A}^{H}_1\mathbf{r}\nonumber\\
\label{hhattheta2}
&=\left(\mathbf{I}_{N_R}\otimes\left((\mathbf{A}^{H}_2\mathbf{A}_2)^{-1}\mathbf
{A } ^ { H
} _2\mathbf{D}^{H}\right)\right)\mathbf{r}.
\end{align}
Using (\ref{A1a}) and (\ref{hhattheta2}), the cost function of the two
dimensional grid search in (\ref{2Dcostfunction}) can be written as,
\begin{align}
 (&\mathbf{r}
-\mathbf{A}_1 \mathbf{A}^{\dag}_1\mathbf{r})^H(\mathbf{r}
-\mathbf{A}_1 \mathbf{A}^{\dag}_1\mathbf{r})\nonumber\\
&=\left(\mathbf{r}
-\left(\mathbf{I}_{N_R}\otimes\left(\mathbf{D}
\mathbf{A}_2(\mathbf{A}^{H}_2\mathbf{A}_2)^{-1}\mathbf{A}^H_2\mathbf{D}
^H\right)\right)\mathbf{r}\right)^H\left(\mathbf{r}
-\left(\mathbf{I}_{N_R}\otimes\left(\mathbf{D}
\mathbf{A}_2(\mathbf{A}^{H}_2\mathbf{A}_2)^{-1}\mathbf{A}^H_2\mathbf{D}
^H\right)\right)\mathbf{r}\right)\nonumber\\
&=\mathbf{r}^H(\mathbf{I}_{N_R}\otimes\mathbf{D})\left(\mathbf{I}_{NN_R}
-\left(\mathbf{I}_{N_R}\otimes
\Pab\right)\right)
\left(\mathbf{I}_{
NN_R}
-\left(\mathbf{I}_{N_R}\otimes
\Pab\right)\right)(\mathbf{I}_{N_R}\otimes\mathbf{D}^H)\mathbf{r}
\nonumber\\
&=\mathbf{d}^T\mathbf{R}^H\left(\mathbf{I}_{NN_R}
-\left(\mathbf{I}_{N_R}\otimes\Pab\right)\right)\left(\mathbf{I}_{
NN_R}
-\left(\mathbf{I}_{N_R}\otimes\Pab\right)\right)\mathbf{R}\mathbf{d^*}
\nonumber\\
\label{dtCChd}
&=\mathbf{d}^T\mathbf{C}\mathbf{C}^H\mathbf{d^*}
\end{align}
\begin{align}
\mathrm{where}\hspace{.2cm}
\mathbf{R}=diag(\mathbf{r}),
\mathbf{d}=diag(\mathbf{I}_{N_R}\otimes\mathbf{D}), \mathbf{D}^H
\mathbf{D}=\mathbf{D}
\mathbf{D}^H=\mathbf{I}_N,
\Pab=\mathbf{A}_2(\mathbf{A}^{H}_2\mathbf{A}_2)^{-1}\mathbf{A}^H_2,\hspace{.1cm}
\mathrm{and}\hspace{.2cm}\nonumber
\end{align}
\begin{align}\label{C}
 \mathbf{C}=\mathbf{R}^H\left(\mathbf{I}_{NN_R}
-\left(\mathbf{I}_{N_R}\otimes\Pab\right)\right).
\end{align}
For small values of $\epsilon$,
$\mathbf{d}\approx\mathbf{1}+j\epsilon\mathbf{c}_1$, where 
 \begin{equation}\label{c1}
\mathbf{c}_1=\displaystyle\frac{2\pi(1+\eta)}{N}diag(\mathbf{I}_{N_R}
\otimes\mathbf{C}_1)
 \end{equation}
with $\mathbf{C}_1$ as given in (\ref{C1}).
Therefore, (\ref{dtCChd}) can be written as,
\begin{align}
 \mathbf{d}^T\mathbf{C}&\mathbf{C}^H\mathbf{d^*}\approx[\mathbf{1}
+j\epsilon\mathbf{ c }_1]^T\mathbf{C}\mathbf{C}^H[\mathbf{1}-j\epsilon\mathbf{
c }_1]\nonumber\\
&=\mathbf{1}^T\mathbf{C}\mathbf{C}^H \mathbf{1}+j\epsilon\mathbf{c}_1^T
\mathbf{C}\mathbf{C}^H \mathbf{1}-j\epsilon\mathbf{1}^T
\mathbf{C}\mathbf{C}^H\mathbf{ c }_1+\epsilon^2 \mathbf{ c }^T_1
\mathbf{C}\mathbf{C}^H \mathbf{ c }_1\nonumber\\
\label{dtCChd_approx}
&=\mathbf{1}^T\mathbf{C}\mathbf{C}^H \mathbf{1}+\epsilon^2\mathbf{ c
}_1^T\Re(\mathbf{C}\mathbf{C}^H)\mathbf{ c
}_1-2\epsilon\mathbf{ c
}_1^T\Im(\mathbf{C}\mathbf{C}^H)\mathbf{1}
\end{align}
Differentiating (\ref{dtCChd_approx}) with respect to $\epsilon$ and equating
to ${0}$ gives the estimate of $\epsilon$ in terms of $\eta$ as, 
\begin{align}\label{eps_hat}
 \hat{\epsilon}=\displaystyle
\frac{\mathbf{c}^T_1\Im(\mathbf{C}\mathbf{C}^H)\mathbf{1}}{\mathbf{c}
^T_1\Re(\mathbf{C}\mathbf{C}^H)\mathbf{c}_1}
\end{align}
Substituting (\ref{eps_hat}) into (\ref{dtCChd_approx}), we have
\begin{align}\label{J3}
 J_3(\eta|\mathbf{r})=\mathbf{1}^T\mathbf{C}\mathbf{C}^H
\mathbf{1}+\hat{\epsilon}^2\mathbf{ c
}_1^T\Re(\mathbf{C}\mathbf{C}^H)\mathbf{ c
}_1-2\hat \epsilon\mathbf{ c
}_1^T\Im(\mathbf{C}\mathbf{C}^H)\mathbf{1}.
\end{align}
Using (\ref{J3}), we get the estimate of $\eta$ by
MML algorithm as 
\begin{align}\label{etahat}
 [\hat{\eta}_{MML}]=\arg\min_{\eta}J_3(\eta|\mathbf{r}).
\end{align}
Substituting (\ref{etahat}) into (\ref{eps_hat}) gives the estimate of
$\epsilon$. Using the estimates of $\epsilon$ and $\eta$, the
estimates of
$\theta$ and $\mathbf{h}$ can be obtained using (\ref{J2}) and
(\ref{hhatML}), as done in Algorithm
$1$. The series of steps involved are given in Algorithm
$2$.
\subsubsection*{\textit{Remarks}}
For the range of values of $|\epsilon|$ less
than $0.10$, the equation 
(\ref{dtCChd_approx}) holds reasonably good with average error of the
approximation less than $10^{-4}$. 
\subsection{Proposed Stage-wise ML (SML) Algorithm}
We also propose SML algorithm
in which the joint estimation of CFO and STE is done in the first
stage using the algorithm in~\cite{joint_FT_CFO1}, ignoring the effect of SFO.
Using the system model in (\ref{datamodel}) and ignoring $\eta$, we have
\begin{align}
\label{AhatSML}
 &\mathbf{A}(\epsilon,0,\theta)=\mathbf{I}_{N_R}\otimes
(\mathbf{D}(\epsilon,0)\mathbf{F}_1(0)\mathbf{G}(\theta)\mathbf{X
(\mathbf{ I } } _ { N_T } \otimes
\mathbf{F_2}))\\
\mathrm{and}\hspace{.5cm}[\hat{\epsilon}_{SML},\hat{\theta}_{SML}]&=\arg\min_{
\epsilon,\theta}(\mathbf{r}
-\mathbf{A} \mathbf{A}^{\dag}\mathbf{r})^H(\mathbf{r}
-\mathbf{A} \mathbf{A}^{\dag}\mathbf{r})\nonumber\\
\label{2Dsearch1}
&=\arg\max_{\epsilon,\theta}\parallel \Pa\mathbf{r}\parallel_{2}^2\\
&=\arg\max_{\epsilon,\theta}J_4(\epsilon,\theta|\mathbf{r})\nonumber\\
\label{J4}
\mathrm{where}\hspace{.2cm}J_4(\epsilon,\eta|\mathbf{r})&=\parallel
\Pa\mathbf{r}\parallel_{2}^2.
\end{align}
With the estimates of CFO and STE from (\ref{2Dsearch1}) in the first stage, the
received signal is used to estimate SFO and channel using the ML algorithm
in~\cite{Gault}, as extended to a MIMO-OFDM system, in the second stage. 
  \begin{align}
 \displaystyle
[\hat{\eta}_{SML}]&=\arg\min_{\eta}(\mathbf{r}
-\mathbf{A} \mathbf{A}^{\dag}\mathbf{r})^H(\mathbf{r}
-\mathbf{A} \mathbf{A}^{\dag}\mathbf{r})\nonumber\\
\label{1Dsearcheta}
&=\arg\max_{\eta}\parallel \Pa\mathbf{r}\parallel_{2}^2\\
&=\arg\max_{\eta}J_5(\eta|\mathbf{r},\hat{\epsilon}_{SML},\hat{\theta}_{SML}
)\nonumber
\end{align}
\begin{equation}\label{J5}
 \mathrm{where}\hspace{.2cm}J_5(\eta|\mathbf{r},\hat{\epsilon}_{SML},\hat{
\theta
}
_{SML})=\parallel
\Pa\mathbf{r}\parallel_{2}^2.
\end{equation}
Finally, using the estimates
of $\epsilon$, $\eta$, and $\theta$, we get the estimate of
$\mathbf{h}$ as in (\ref{hhatML}).
The series of steps involved are given in Algorithm $3$.
\begin{algorithm}
\label{MLAlgo}
\caption{SML Algorithm}
{\bf Inputs:} $N$, $N_T$, $N_R$, $\mathbf{r}$, $\mathbf{X}$,
$L_m$, $[\theta_{min}, \theta_{max}, \theta_{grid}]$, $[\epsilon_{min},
\epsilon_{max},  \epsilon_{grid}]$, $[\eta_{min}, \eta_{max}, \eta_{grid}]$
\label{OMPAlgo}
 \begin{algorithmic}[1]
   \STATE Evaluate $\mathbf{F}_2$, $\mathbf{F}_1(0)$;\hfill $\star$
using (\ref{F_2}), (\ref{F_1}) and $\eta$ is ignored

      \FOR{$i=$ $\epsilon_{min}:\epsilon_{grid}:\epsilon_{max}$}
      \FOR{$k=$ $\theta_{min}:\theta_{grid}:\theta_{max}$}
	\STATE Evaluate $\mathbf{D}(i,0)$, $\mathbf{G}(k)$;\hfill $\star$
using (\ref{D}), (\ref{G}) and $\eta$ is ignored
\STATE $\mathbf{A}(i,0,k)=\mathbf{I}_{N_R}\otimes
\left(\mathbf{D}(i,0)\mathbf{F}_1(0)\mathbf{G}(k)\mathbf{X}(\mathbf{I}_{N_T}
\otimes
\mathbf{F}_{2})\right)$
;\hfill $\star$ using (\ref{AhatSML})
\STATE $J_4(i,k|\mathbf{r})=\parallel \Pa\mathbf{r}\parallel_{2}^2$;\hfill
$\star$ using (\ref{J4})
      \ENDFOR
    \ENDFOR
  \STATE $\displaystyle
[\hat{\epsilon}_{SML},\hat{\theta}_{SML}]=\arg\max_{i,k}J_4(i,k|\mathbf{r})$;
\STATE Evaluate $\mathbf{G}(\hat{\theta}_{SML})$;\hfill $\star$
using (\ref{G})
\FOR{$j=$ $\eta_{min}:\eta_{grid}:\eta_{max}$}
\STATE Evaluate $\mathbf{D}(\hat{\epsilon}_{SML},j)$, $\mathbf{F}_1(j)$,
$\mathbf{A}(\hat{\epsilon}_{SML},j,\hat{\theta}_{SML})$;\hfill
$\star$ using (\ref{D}), (\ref{F_1}), (\ref{A})
\STATE $J_5(j|\mathbf{r},\hat{\epsilon}_{SML
},\hat{\theta}_{SML})=\parallel \Pa\mathbf{r}\parallel_{2}^2$;\hfill
$\star$ using (\ref{J5}) 
\ENDFOR
\STATE $\displaystyle
[\hat{\eta}_{SML}]=\arg\max_{k}J_5(j|\mathbf{r},\hat{\epsilon}_{SML
},\hat{\theta}_{SML})$;
\STATE
$\hat{\mathbf{A}}(\hat{\epsilon}_{SML},\hat{\eta}_{SML},\hat{\theta}_{SML}
)$;\hfill $\star$ using (\ref{A})
\STATE
$[\hat{\mathbf{h}}_{SML}]=\hat
{\mathbf{A}}^\dag\mathbf{r}$;\hfill
$\star$ using (\ref{hhatML}) 
\end{algorithmic}
{\bf Output:}
$[\hat{\theta}_{SML},
\hat{\epsilon}_{SML}, \hat{\eta}_{SML}, \hat{\mathbf{h}}_{SML}]$ 
 \end{algorithm}
\subsection{Comparison of Computational Complexity}
The computational complexity of ML algorithm depends mainly on 
computation of the two-dimensional grid search in (\ref{2Dsearch}), for
obtaining the estimates of $\epsilon$ and $\eta$, and the one-dimensional
 grid search in (\ref{1Dsearch}), for obtaining the estimate of $\theta$. Let
$g_{\epsilon}$, $g_{\eta}$, and $g_{\theta}$ denote the number of grid points
used in search for $\epsilon$, $\eta$, and $\theta$. Therefore, the
computational complexity of evaluating (\ref{2Dsearch}) is approximately equal
to $g_{\epsilon}g_{\eta}\mathcal{O}(N^3)$, whereas that of evaluating
(\ref{1Dsearch}) is approximately equal to 
 $g_{\theta}\mathcal{O}(N^3)$. 
Thus, the total computational complexity of
ML algorithm is approximately given by
 $(g_{\epsilon}g_{\eta}+g_{\theta})\mathcal{O}(N^3)$. 
\par
The computational complexity of MML algorithm depends mainly on the 
computation of two one-dimensional grid searches, given in
(\ref{etahat}) and (\ref{1Dsearch}). The
computational complexity of evaluating (\ref{etahat}) is approximately equal
to $g_{\eta}\mathcal{O}(N^3)$, whereas that of evaluating
(\ref{1Dsearch}) is approximately equal to 
$g_{\theta}\mathcal{O}(N^3)$. Thus, the total computational complexity of
MML algorithm is approximately given by
$(g_{\eta}+g_{\theta})\mathcal{O}(N^3)$.
\par
The computational complexity of SML algorithm depends mainly on the 
computation of the two-dimensional grid search in (\ref{2Dsearch1}), for the
estimation 
of $\epsilon$ and $\theta$ in the first stage, and the 
 one-dimensional grid search in (\ref{1Dsearcheta}), for the estimation of
$\eta$ in the second stage. The
total computational complexity of the first stage is approximately equal
to $g_{\epsilon}g_{\theta}\mathcal{O}(N^3)$, whereas that of second stage is
approximately equal to 
$g_{\eta}\mathcal{O}(N^3)$. Thus, the total computational complexity of
SML algorithm is approximately given by
$(g_{\epsilon}g_{\theta}+g_{\eta})\mathcal{O}(N^3)$.
\subsubsection*{Remarks}
Based on the system setup in Section V, the typical values of $g_{\epsilon}$,
$g_{\eta}$, and $g_{\theta}$ are $81$, $101$, and $43$, respectively. Thus, MML
algorithm is around 57 times and 25 times faster than ML and SML
algorithms, respectively.
\section{Simulation Results and Discussions}\label{simulation}
\subsection{System Setup}
 The simulated $2\times 2$ MIMO-OFDM system has $N=128$ subcarriers for each
transmitter with $20$ MHz signal
bandwidth.
 The SCM specified by 3GPP~\cite{SCM1}, is used to
generate the fading channels
in the Urban Macro Scenario with $L_m=10$. Also, The
transmitted symbols belong to QPSK constellation with unit amplitude. Therefore,
the variance of the complex additive Gaussian noise at the
receiver, $\sigma^{2}_{w}$, is
varied for getting SNR(dB) in steps of $5$ for the evaluation of Mean Square
Error (MSE) and CRLB. Also, $10^3$ trials are carried out in simulations for
getting each point in the MSE and CRLB plots. 
\par
We consider the training blocks having a CP of length $32$. The condition
$(L_m+\theta_{max})$ less
than length of CP results in $\theta_{max}=21$ and $|\theta|<21$.
 The range of normalized CFO used for grid search is
$|\epsilon|<0.4$ with a resolution of $10^{-2}$ and that of normalized SFO
 is $|\eta|<5\times10^{-3}$ with a resolution of $10^{-4}$.
The actual values of the impairments, $\epsilon$, $\eta$, and $\theta$ used in
the simulations are $0.021$, $0.000101$ or $101$ $\mathrm{ppm}$, and $2$,
respectively.  
  \subsection{Performance Measures}
The ML estimates of the parameters are used for calculating
the MSE values as,
\begin{equation}\label{mse}
 \mathrm{MSE}(\hat{\boldsymbol{\rho}})=\frac{\sum\limits_{i=1}^{N_{trials}}
\parallel\hat{\boldsymbol
{\rho}}_{i}-\boldsymbol{\rho}\parallel^{2}_2}{N_{trials}}.
\end{equation}
where $\boldsymbol{\rho}$ represents the actual
parameter, $\hat{\boldsymbol{\rho}}_{i}$
represents the estimate of the
parameter at
$i^\mathrm{th}$ trial, and $N_{trials}$ represents the number of trials.
The Probability of Timing Failure $(P_{tf})$ is used as an indicator to
illustrate the robustness of symbol timing
estimator~\cite{Saemi},
 which is
expressed as,
\begin{equation}\label{Ptf1}
 P_{tf}(p)=Pr\left[|\hat{\theta}-\theta|\geqslant p\right]
\end{equation}
where $p$ is the absolute difference between the estimated value
of $\theta$ and
 the actual value of $\theta$.
\subsection{Performance Assessment}
\begin{figure}[]
\centering{
\resizebox{9.5cm}{!}{
\includegraphics[]{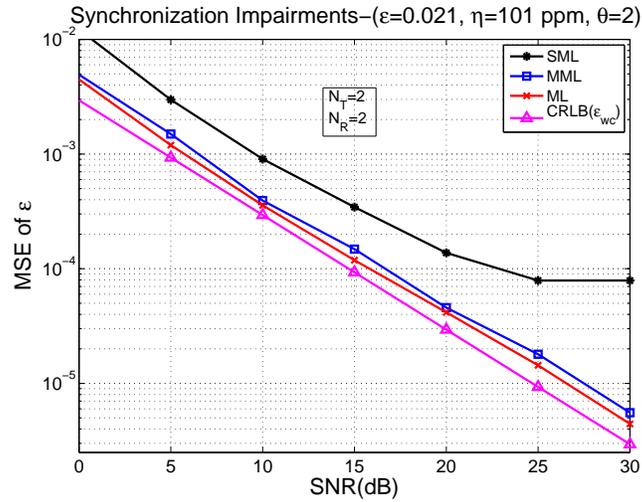}}}
\caption{CRLB and MSE for the estimation of CFO as a function of
SNR(dB) using ML, MML, and SML
algorithms.}
\label{MSECFO}
\end{figure}

 The MSE values of the estimated parameters
  are calculated using
(\ref{mse}) and are plotted in log-scale against SNR(dB) for ML, MML, and SML
algorithms in Fig.$2$-Fig.$5$.
 The CRLBs of the parameters are also plotted in the corresponding figures.
\begin{figure}[]
\centering{
\resizebox{9.5cm}{!}{
\includegraphics[]{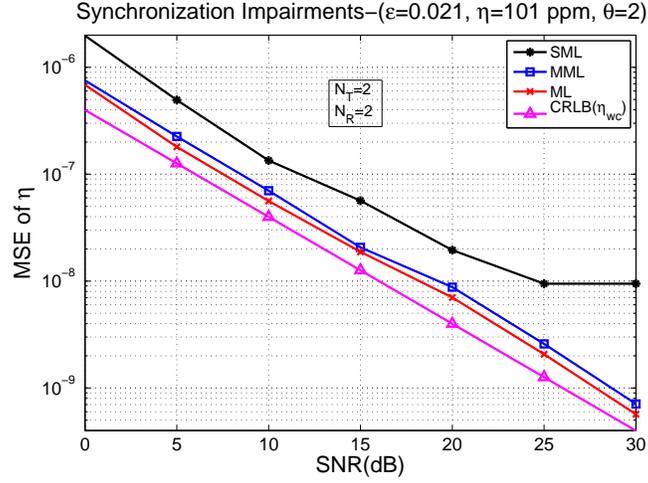}}}
\caption{CRLB and MSE for the estimation of SFO as a function of
SNR(dB) using ML, MML, and SML
algorithms.}
\label{MSEETA}
\end{figure}
It is found from Fig.\ref{MSEETA} and Fig.\ref{MSECFO} that the MSE plots of
the ML algorithm for the estimation of $\eta$ and $\epsilon$ closely follows
$\mathrm{CRLB}(\eta_{wc})$ and $\mathrm{CRLB}(\epsilon_{wc})$, but with a
performance degradation of around $1.25$ dB SNR and $0.75$ dB SNR,
respectively. Also, it is found that there is only small performance difference
between ML and MML algorithms.  
 Performance loss of more than $3$dB, and error floor at $25$ dB occurs for SML
algorithm when compared with the performance of ML or MML algorithms in
Fig.\ref{MSEETA} and Fig.\ref{MSECFO}, respectively. 
\begin{figure}[]
\centering{
\resizebox{9.5cm}{!}{
\includegraphics[]{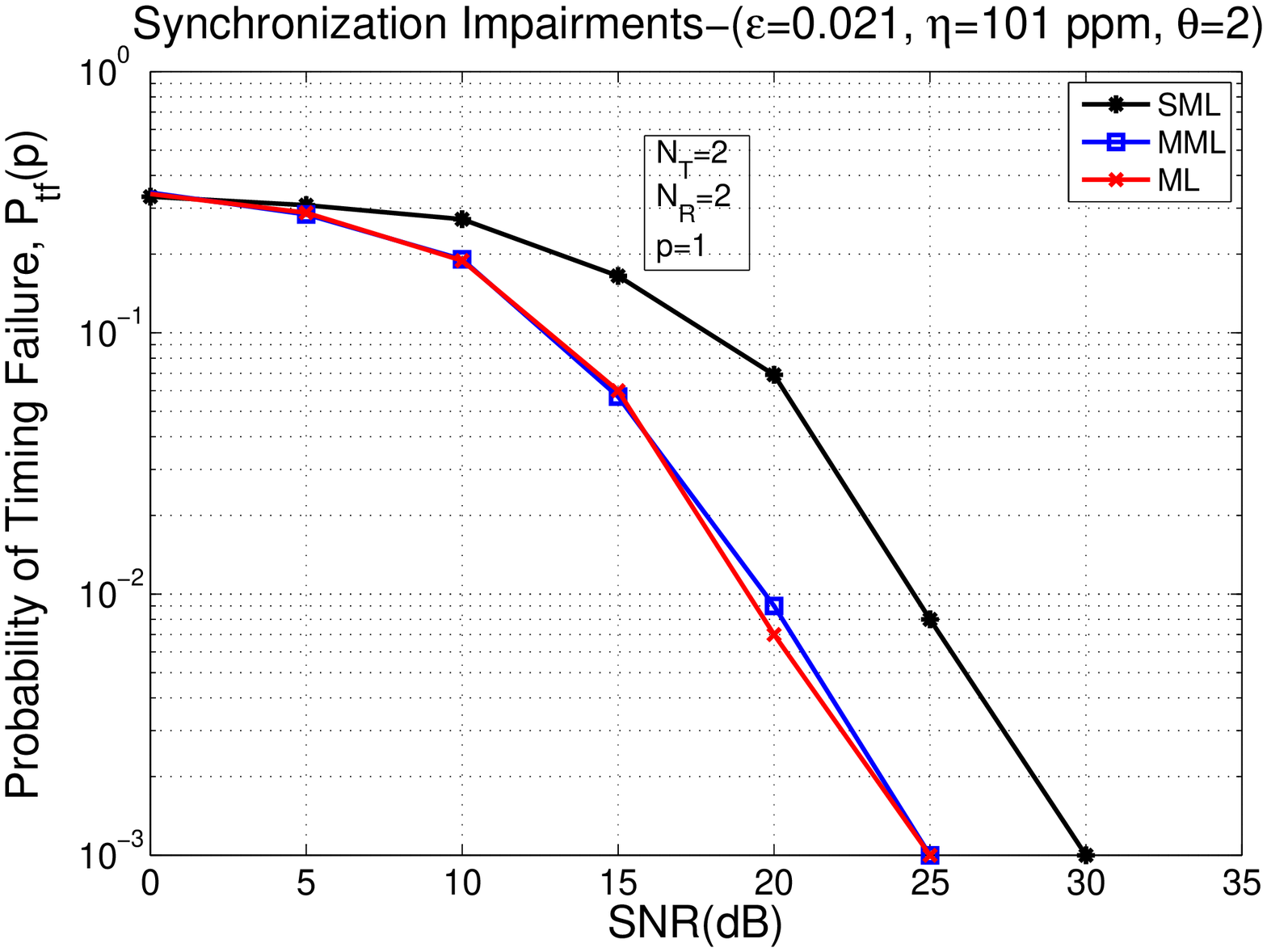}}}
\caption{Probability of Timing Failure as a function of SNR(dB) using ML, MML,
and SML algorithms.}
\label{Ptf}
\end{figure}
\par
The probability of timing failure for the estimation of $\theta$ is calculated
for $p=1$, as defined in (\ref{Ptf1}), and is plotted in Fig.\ref{Ptf} for
 ML, MML, and SML algorithms, respectively.  
As in the cases of $\epsilon$ and $\eta$, there is only small
performance degradation due to the approximation in the MML
algorithm, and performance loss of more than $3$dB occurs for SML algorithm
when compared with ML algorithm, as observed from the plots
 in Fig.\ref{Ptf}.  
\begin{figure}[htb]
\centering{
\resizebox{9.5cm}{!}{
\includegraphics[]{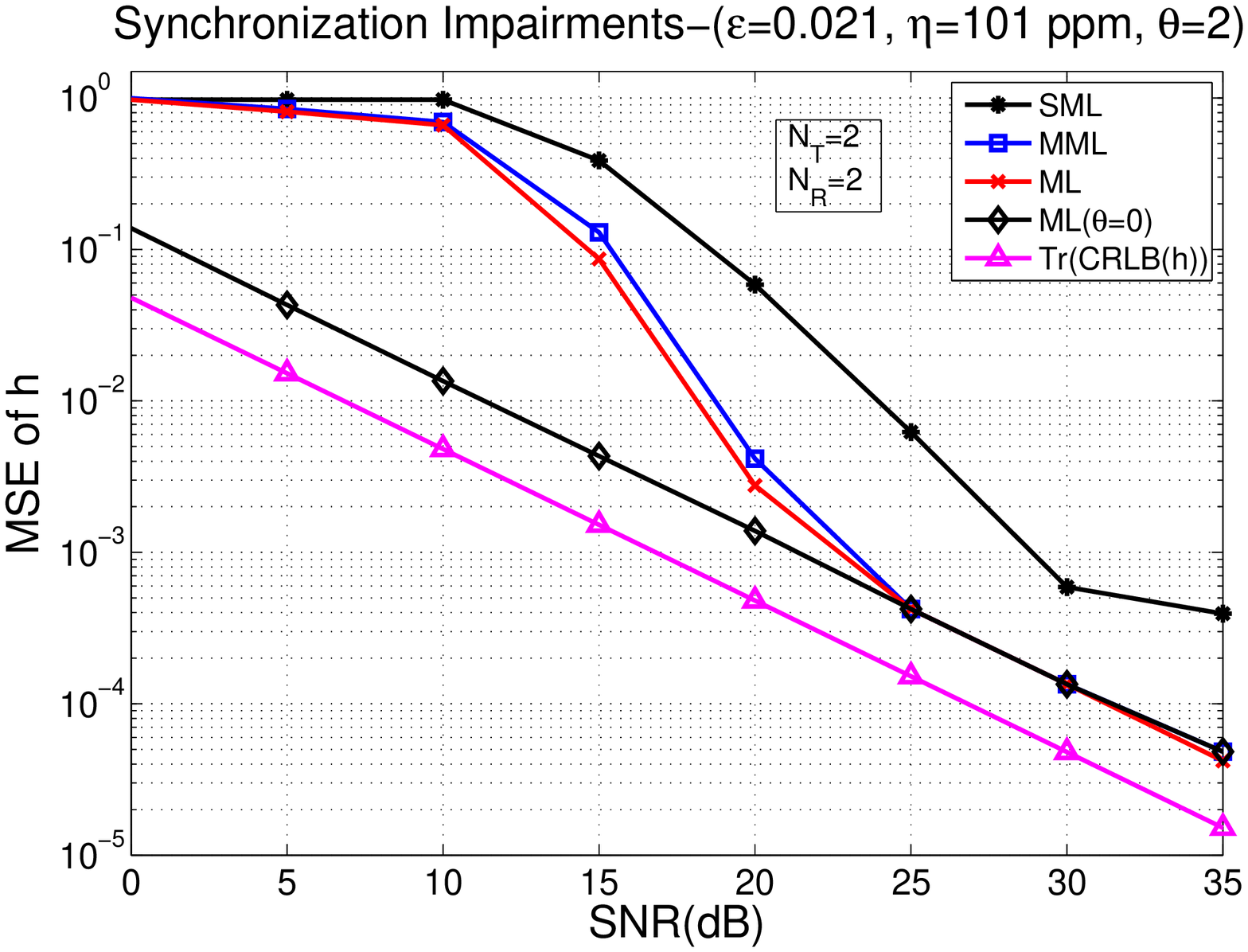}}}
\caption{CRLB and MSE for the estimation of channel as a function of SNR(dB)
using ML, MML, and SML algorithms.}%
\label{MSEh}
\end{figure}
The MSE plots of ML estimates of
$\mathbf{h}$ using ML algorithm with STE=0, denoted as ML($\theta=0$),
together with ML, MML, and SML algorithms are shown in Fig.\ref{MSEh}.
 Tr($\mathrm{CRLB}$($\mathbf{h}$))
evaluated using (\ref{hwc}) is also shown in the same
figure. 
It is found from Fig.\ref{MSEh} that
MSE plot of ML($\theta=0$) follows
Tr($\mathrm{CRLB}(\mathbf{h})$) with a performance loss of $4$ dB. Comparing the
MSE plots of ML($\theta=0$) and ML in the figure, it is found
that the
performance degradation of ML algorithm below $20$ dB SNR is due to
the coupling effect of STE on the channel. As in the cases of $\epsilon$, $\eta$
and $\theta$, there is only small
performance degradation due to the approximation in the MML
algorithm.
  Also, the performance degradation and
error floor of the MSE plot of SML 
algorithm is due to the stage-wise estimation of the parameters.

Thus, it is
found from the figures that there is only
small performance degradation due to the approximation in the MML
algorithm when compared with ML algorithm. Also, performance
degradation occurs, if joint estimation of all impairments together is not done,
as observed from the performance plots of
 SML algorithm in the figures.  
\section{Conclusion}\label{Conclusion}
 In this paper, ML algorithms, for the joint estimation of synchronization
impairments and channel in a MIMO-OFDM system, are proposed.
 A system model, which shows the effects of the synchronization impairments and
channel, is formulated. 
CRLBs for the estimation of continuous parameters are derived, which show the
coupling effect
among different parameters and the significance of a joint estimator.  
An ML estimation algorithm is proposed, where the multi-dimensional
optimization problem for estimating the parameters is reduced to a
two-dimensional and a one-dimensional grid search.
To reduce the complexity further, MML algorithm 
which involves only multiple one-dimensional searches is also proposed.
 Further, SML algorithm using existing algorithms, which estimate
fewer number of parameters, is also proposed.
The performances of the estimation methods are studied through numerical
simulations and it is observed that, there is only small performance degradation
due to the approximation in MML algorithm. Also, the proposed ML and MML
algorithms exhibit
better performance than the proposed SML algorithm, which uses existing
algorithms. 
\bibliography{IEEEabrv,reference}

\begin{thebibliography}{10}

\bibitem{Meyr}
M.~Speth, S.~Fechtel, G.~Fock, and H.~Meyr, ``Optimum receiver design for
  wireless broad-band systems using ofdm. 1,'' {\em {IEEE} Trans. Commun.},
  vol.~47, pp.~1668 --1677, Nov 1999.

\bibitem{morelli_tutorial}
M.~Morelli, C.-C. Kuo, and M.-O. Pun, ``Synchronization techniques for
  orthogonal frequency division multiple access (ofdma): A tutorial review,''
  {\em Proceedings of the IEEE}, vol.~95, pp.~1394 --1427, july 2007.

\bibitem{morelli2}
M.-O. Pun, M.~Morelli, and C.-C. Kuo, ``Maximum-likelihood synchronization and
  channel estimation for ofdma uplink transmissions,'' {\em {IEEE} Trans.
  Commun.}, vol.~54, pp.~726 -- 736, april 2006.

\bibitem{Nugen}
H.~Nguyen-Le, T.~Le-Ngoc, and C.~C. Ko, ``Joint channel estimation and
  synchronization for mimo ofdm in the presence of carrier and sampling
  frequency offsets,'' {\em {IEEE} Trans. Vehicular Technol.}, vol.~58,
  pp.~3075 --3081, july 2009.

\bibitem{joint_FT_CFO}
J.-W. Choi, J.~Lee, Q.~Zhao, and H.-L. Lou, ``Joint ml estimation of frame
  timing and carrier frequency offset for ofdm systems employing time-domain
  repeated preamble,'' {\em {IEEE} Trans. Wireless Commun.}, vol.~9, pp.~311
  --317, january 2010.

\bibitem{joint_FT_CFO1}
S.~Salari and M.~Heydarzadeh, ``Joint maximum-likelihood estimation of
  frequency offset and channel coefficients in multiple-input multiple-output
  orthogonal frequency-division multiplexing systems with timing ambiguity,''
  {\em IET Commun.}, vol.~5, pp.~1964 --1970, 23 2011.

\bibitem{Saemi}
A.~Saemi, V.~Meghdadi, J.-P. Cances, and M.~Zahabi, ``Joint ml time-frequency
  synchronisation and channel estimation algorithm for mimo-ofdm systems,''
  {\em IET Circuits Devices Syst.}, vol.~2, pp.~103 --111, february 2008.

\bibitem{Gault}
S.~Gault, W.~Hachem, and P.~Ciblat, ``Joint sampling clock offset and channel
  estimation for ofdm signals: Cramer-rao bound and algorithms,'' {\em {IEEE}
  Trans. Signal Process.}, vol.~54, pp.~1875 -- 1885, may 2006.

\bibitem{KimML}
Y.-H. Kim and J.-H. Lee, ``Joint maximum likelihood estimation of carrier and
  sampling frequency offsets for ofdm systems,'' {\em {IEEE} Trans.
  Broadcast.}, vol.~57, pp.~277 --283, june 2011.

\bibitem{renu}
R.~Jose and K.~V.~S. Hari, ``Joint estimation of synchronization impairments in
  mimo-ofdm system,'' in {\em 2012 National Conference on Communications
  (NCC)}, pp.~1 --5, feb. 2012.

\bibitem{startofframe1}
T.~Schmidl and D.~Cox, ``Robust frequency and timing synchronization for
  ofdm,'' {\em {IEEE} Trans. Commun.}, vol.~45, pp.~1613 --1621, dec 1997.

\bibitem{startofframe2}
H.~Minn, V.~Bhargava, and K.~Letaief, ``A robust timing and frequency
  synchronization for ofdm systems,'' {\em {IEEE} Trans. Wireless Commun.},
  vol.~2, pp.~822 -- 839, july 2003.

\bibitem{startofframe3}
K.~Shi and E.~Serpedin, ``Coarse frame and carrier synchronization of ofdm
  systems: a new metric and comparison,'' {\em {IEEE} Trans. Wireless Commun.},
  vol.~3, pp.~1271 -- 1284, july 2004.

\bibitem{morelli3}
M.~Morelli, ``Timing and frequency synchronization for the uplink of an ofdma
  system,'' {\em {IEEE} Trans. Commun.}, vol.~52, p.~166, jan. 2004.

\bibitem{timing_fraction}
T.~Lv, H.~Li, and J.~Chen, ``Joint estimation of symbol timing and carrier
  frequency offset of ofdm signals over fast time-varying multipath channels,''
  {\em {IEEE} Trans. Signal Process.}, vol.~53, pp.~4526 -- 4535, dec. 2005.

\bibitem{stoica}
P.~Stoica and O.~Besson, ``Training sequence design for frequency offset and
  frequency-selective channel estimation,'' {\em {IEEE} Trans. Commun.},
  vol.~51, pp.~1910 -- 1917, nov. 2003.

\bibitem{morelli4}
M.~Morelli and U.~Mengali, ``Carrier-frequency estimation for transmissions
  over selective channels,'' {\em {IEEE} Trans. Commun.}, vol.~48, pp.~1580
  --1589, sep 2000.

\bibitem{NugenRLS}
H.~Nguyen-Le, T.~Le-Ngoc, and C.~C. Ko, ``Rls-based joint estimation and
  tracking of channel response, sampling, and carrier frequency offsets for
  ofdm,'' {\em {IEEE} Trans. Broadcast.}, vol.~55, pp.~84 --94, march 2009.

\bibitem{Kay}
S.~M. Kay, {\em Fundamentals of statistical signal processing: estimation
  theory}.
\newblock Upper Saddle River, NJ, USA: Prentice-Hall, Inc., 1993.

\bibitem{Horn_Johnson}
R.~A. Horn and C.~R. Johnson, {\em Matrix Analysis}.
\newblock Cambridge University Press, 1990.

\bibitem{SCM1}
``Spatial channel model for multiple input multiple output (mimo)
  simulations,'' {\em 3GPP}, vol.~TR 25.996, no.~v6.1.0, Sep. 2003.

\end{thebibliography}
\bibliographystyle{ieeetr}

\end{document}